\documentclass[12pt,preprint]{aastex} 

\usepackage{color,epsfig} 
 
\shorttitle{Carbon Monoxide in Cas A} 
\shortauthors{}

\newcommand{\spitzer}{\textit{Spitzer}} 

\newcommand{\mic}{$\mu$m} 

\begin{document} 
\title{Carbon Monoxide in the Cassiopeia A Supernova Remnant} 
 
\author{ 
J. Rho\altaffilmark{1},  
T. H. Jarrett\altaffilmark{1}, 
W. T. Reach\altaffilmark{1},
H. Gomez\altaffilmark{2}, and
M. Andersen\altaffilmark{1},
} 
\altaffiltext{1}{{\it Spitzer} Science Center, California 
Institute of Technology,  Pasadena, CA 91125; rho, jarrett, reach, mortena@ipac.caltech.edu} 
\altaffiltext{2}{School of Physics and Astronomy, Cardiff University, The Parade, Cardiff, CF24 3AA, UK;
gomez@astro.cf.ac.uk} 
 
\begin{abstract}

 We report the likely detection of  near-infrared 2.29 $\mu$m first
overtone Carbon Monoxide (CO) emission from the young supernova remnant
Cassiopeia A (Cas A).   The continuum-subtracted CO filter map reveals
CO knots within the ejecta-rich  reverse shock.  We compare the first
overtone CO emission with that found in the well-studied supernova, SN
1987A and find $\sim$30 times less CO in Cas A.  The presence of CO
suggests that molecule mixing is small in the SN ejecta and that
astrochemical processes and molecule formation may continue at least
$\sim 300$ years after the initial explosion.


\end{abstract} 
\section{Introduction} 
Supernovae (SNe) are suggested to be the first molecular factories in 
the early Universe \citep{cherchneff08}. Molecules such as carbon 
monoxide (CO) are sensitive diagnostics of  the temperature, density,
and degree of mixing  in the SN ejecta.  Understanding the chemistry
within the ejecta, and the abundance and mixing of CO and SiO
molecules, is particularly important in studying dust formation
processes since this controls how much carbon is available to form
amorphous C grains and how much  oxygen available beyond that bound in
CO and SiO molecules to form MgSiO$_3$ and  Mg$_2$SiO$_4$.  The
existence of CO in supernova ejecta suggests that  dust formation can
occur, since molecules are indicative of  relatively cool ejecta
(typically few thousand degrees).   CO itself is  an effective coolant
\citep{liu95}, so that once it forms it could enhance dust formation. 
Isotopic compositions of  meteoric dust grains \citep{clayton04} show
that carbon-rich dust is  condensed within the interior of supernova
ejecta.   In O-rich ejecta,  radioactivity and mixing processes produce
`free' carbon in which to  form dust. 

 The {\em quantity} of dust formed within SNe ejecta has become a
subject of debate in recent years (see Rho et al. 2008 and references
therein; Dunne et al. 2008).  \citet{lagage96} first showed evidence,
in {\it Infrared Space Observatory} ({\it ISO}) images, for dust
formation in the nearby, young supernova remnant (SNR) Cas A.
\cite{rho08} identified three distinct ejecta dust populations  based
on their continuum shapes in {\it Spitzer Space Telescope} spectra, 
including broad features peaking at  21$\mu$m  and estimated a total
dust mass of between 0.02 to $\rm  0.054\,\rm M_\odot$.  These
observations provide definitive evidence of  freshly-formed dust in a
young supernova remnant and  show that SNe are  important dust
contributors to the large  quantities of dust observed at high
redshifts. Theoretical models \citep{todini01, nozawa03, bianchi07}
predict a dust mass of 0.3-1 M$_{\odot}$ per SN. However, our current
understanding of dust chemistry and composition is limited. The current
models assume that very little carbon is locked up in CO and almost all
of carbon is available for dust formation.  The results from dust
evolution models therefore depend significantly on the C/O ratio in the
gas  (see Morgan \& Edmunds 2003 and Dwek et al. 1998).

Images obtained with the \spitzer\ Infrared Array Camera (IRAC) each
present a different view of the nucleosynthesis within Cas A
\citep{ennis06}. A surprising result from the IRAC observations is the
similarity between the morphology of the IRAC Band 2 image (centered at 4.5
$\mu$m) and ejecta-dominated images of the remnant.  The origin
of this emission is not known, but based on the lack of bright
H$\alpha$ emission, the contribution from Br$\alpha$ emission is
likely small. \cite{ennis06} therefore suspected that CO emission was
responsible for the emission at this wavelength, but could not be
confirmed since the {\it Spitzer} Infrared Spectrograph (IRS)
has a blue cutoff of 5.2 \mic.

As a tracer of gas properties and dust formation, CO is potentially an 
important diagnostic for any supernova.  The first detection of CO 
from supernova ejecta was from the well-studied, Type II supernova,
SN1987A  \citep{catchpole89, meikle89, spyromilio88}, providing an
unique  insight into the chemical dynamics of the explosion and
ejecta.  The variation of the observed CO emission with time after the
inital explosion was modeled by \cite{spyromilio88}, who suggested that
CO formed out in the SN ejecta at a temperature of 1800\,K and
expansion velocity 1200 km~s$^{-1}$ at 337 days after the explosion. 
They determined a total mass  $\rm M_{CO} \sim 10^{-3} M_{\odot}$. 
\cite{spyro96} detected $\rm \sim 10^{-3} \,M_{\odot}$ of CO forming at
temperature 4000\,K in SN 1995ad.  CO has further been detected from
the Type IIn SN 1998S \citep{gerardy00}, SN 2002dh \citep{pozzo06}, and
the Type Ic SN 2000ew \citep{gerardy02}. The number of CO  detections
in SNe in the literature is still relatively small, and all are 
restricted to within a few years after the initial explosion.  In this 
{\it Letter}, we report the likely detection of the CO overtone from
Cas  A, an approximately 300-yr old supernova remnant. This is  the
first time CO emission has been detected in a Galactic supernova
remnant  and suggests that molecule formation is not only a common 
occurrence in core-collapse ejecta but molecules also exist (or
continue  forming) long after the initial explosion event.  
 
\section{Observations} 
 
We observed Cas A on 2006, September 3 and 4 with the Wide field
InfraRed Camera (WIRC, Wilson et al. 2003) on the Hale 200 inch (5 m)
telescope at Mount Palomar.  The weather was very clear with the seeing
typically 0.8$''$.  The camera field of view is $8.7'$. We observed a
set of dithered positions (designed to uniformly cover the supernova
remnant), with interleaved, dithered observations of an `off' position
$30'$ away. A sky reference image was generated from the `off' images,
and was subtracted from the `on' images after scaling to match the
median sky brightness. This observation and analysis procedure
preserved the diffuse emission of the SNR.  We took narrow-band images
using filters centered on CO(2-0) at 2.294 \mic, K-band and H-band
continuum at 2.270 and 1.570 \mic, and P$\beta$ at 1.182 \mic\ (see
Figure \ref{casacothreecolor}).  The narrow-band filter widths are
typically 0.025 \mic. The exposure time of CO, K-cont and P$\beta$ is a
combined 90 sec from three back-to-back images of 30 sec exposure (to
reduce the number of saturated stars), with a total on-source
integration time of 8100 sec for CO, 7290 sec for K-cont and 810 sec
for P$\beta$ image.  The exposure time of H-cont is 90 sec, with a
total on-source integration time of 1440 sec.  Respective sensitivities
of the narrow-filter images are 1.70$\times$10$^{-6}$ $\rm erg~
s^{-1}~cm^{-2} ~sr^{-1}$ for CO, 1.04$\times$10$^{-6}$ for K-cont,
1.54$\times$10$^{-6}$ for H-cont, and 1.25$\times$10$^{-5}$ for
P$\beta$.  The detailed data reduction methods are similar to those
described in \cite{keohane07}; additionally, focal plane distortion was
derived from the images and corrected.  We directly subtracted the
K-continuum image from the CO filter image in order to subtract stars
and contamination from synchrotron emission from Cas A \citep{rho03}.
Due to excellent seeing (0.8$''$) and low airmass towards Cas A, the
stars were well subtracted. We further masked saturated stars and
applied adaptive smoothing and wavelet filtering to the images.
 
\section{Results} 
\begin{figure} 
\epsscale{0.94}
\plotone{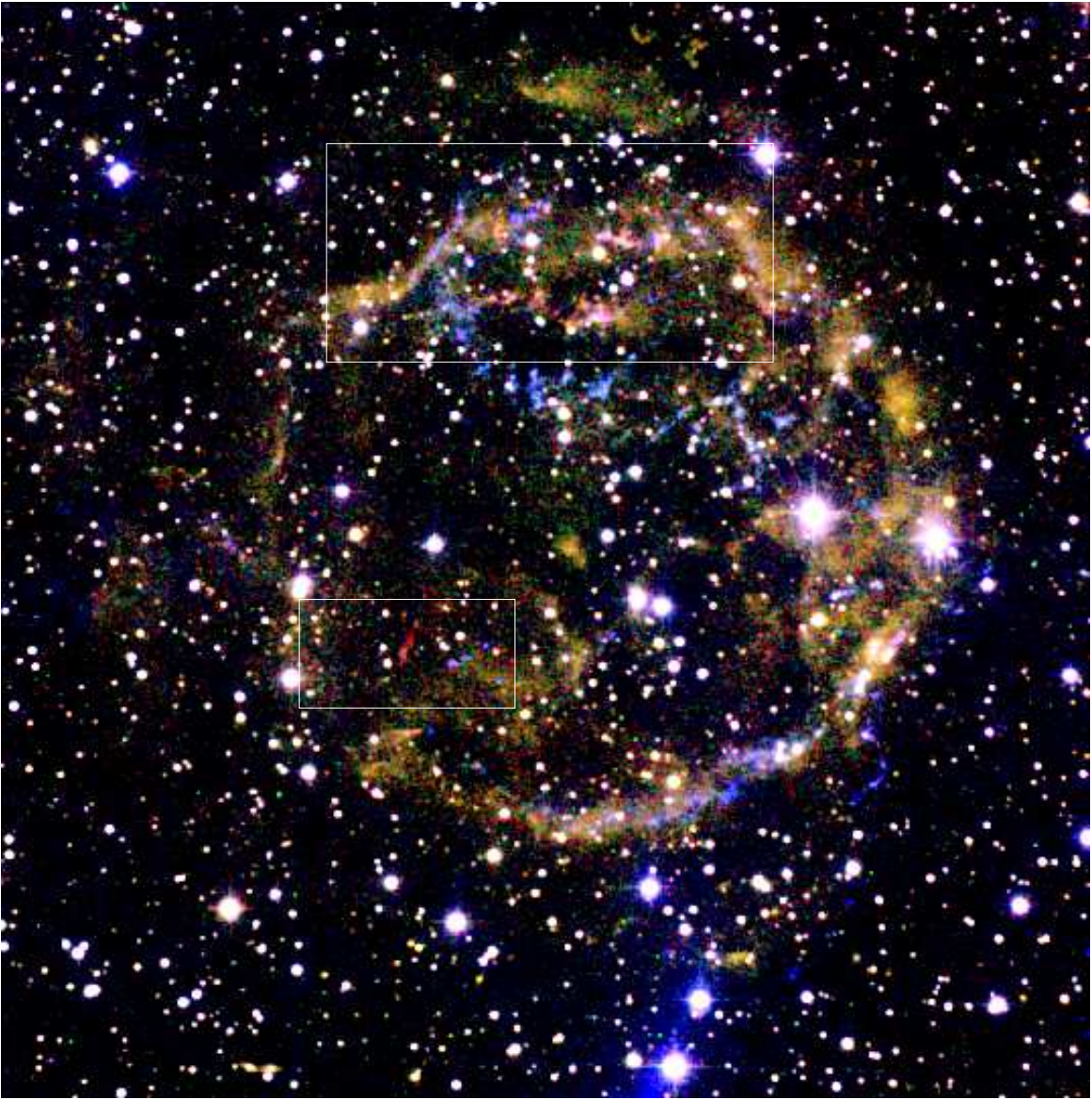}
\caption{Mosaicked color image of Cas A with WIRC, with CO (red),
K-continuum (green) and P$\beta$ (blue).   The range of surface
brightnesses measured from the diffuse structures are 0.31 - 2.79$\times$10$^{-5}$ for CO,  0.23 - 1.4$\times10^{-5}$ for K-continuum, and 1.2 - 29.0$\times10^{-5}$ erg
s$^{-1}$ cm$^{-2}$ sr$^{-1}$  for P$\beta$. CO-excess regions (in
red) are marked as boxes (the bottom box marks the location of the Minkowski knot). The image is centered at R.A.\ $350.858^\circ$ and Dec.\ $58.814^\circ$ (J2000), and covers  an 7$'$$\times$7$'$ arcmin
field of view. }
\label{casacothreecolor} 
\end{figure}

\begin{figure} 
\epsscale{0.92} 
\plotone{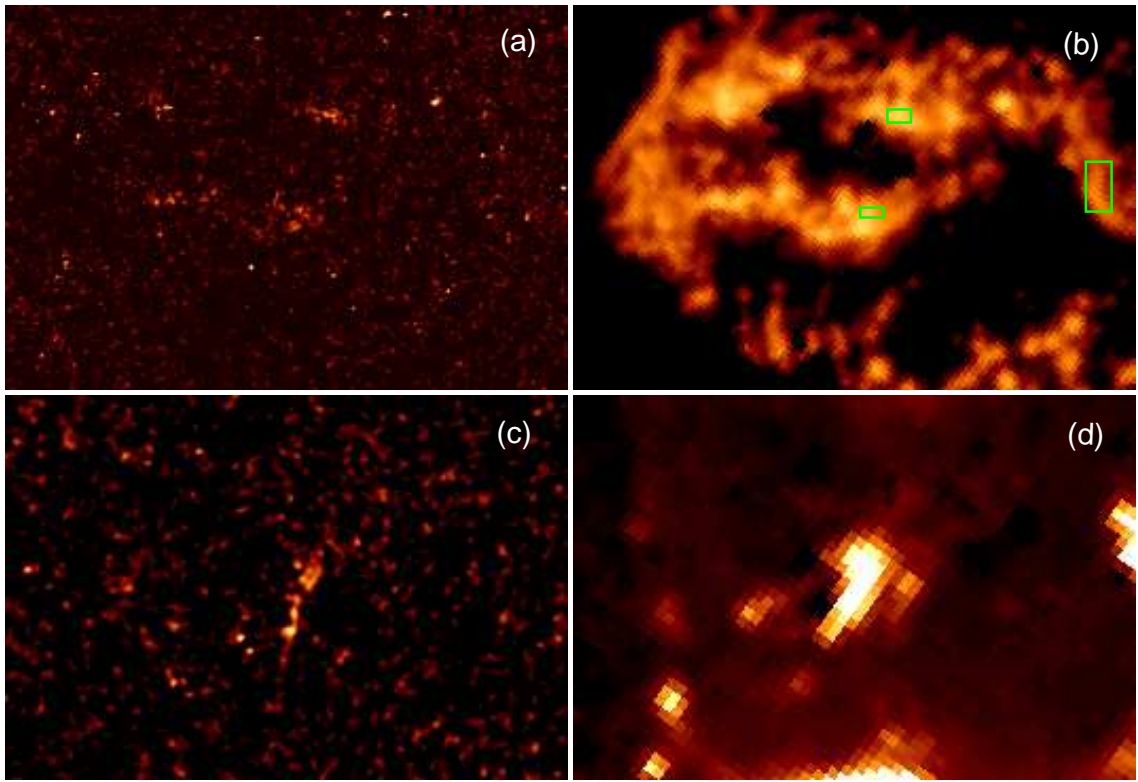}
\caption{Zoomed-in continuum-subtracted-CO (a and c) and IRAC2 (b and d)
color images  towards CO-detected regions.
These zoomed-in images
 cover the regions marked
as boxes in Figure \ref{casacothreecolor}: the pannels (a) and (b) show the
northern bright ring  [centered at R.A.\ $350.857^\circ$ and Dec.\
$58.836^\circ$ (J2000), with area 2.5$'$$\times$1.2$'$], and pannels (c)
and (d) are  towards the Minkowski knots [centered at R.A.\
$350.888^\circ$ and Dec.\ $58.802^\circ$ (J2000), with area
1.2$'$$\times$0.6$'$ ], respectively. The near-infrared CO images (a and
c) include the CO (2-0) first overtone, and the {\it Spitzer} IRAC 2
images (b and d) include CO (1-0) fundamental.  The locations of the
three knots  in Table 1 are marked as green rectangles  (upper: CO knot1, lower: CO
knot2, and right: NW knot).
}
\label{casacoonly} 
\end{figure}

Figure \ref{casacothreecolor} shows a color composite image of Cas A,
combining the P$\beta$ line (blue), 2.27 \mic\ continuum (green), and
CO (2-0) 2.29 \mic\ line (red).    P$\beta$ traces recombining atomic
hydrogen. The 2.27 \mic\ continuum consists mostly of non-thermal
synchrotron radiation and delineates the shell structure (see Rho et
al.  2003). Finally, the CO (2-0) overtone emission traces the dense
knots of gas along the shell and interior interface. The molecular
emission is evident as red in Figure \ref{casacothreecolor} and more
easily seen in the continuum-subtracted CO image,  Figure
\ref{casacoonly}, revealing the line emission towards the northern
parts of bright rings  and an eastern filament of the Minkowski knots
in the southeast. The CO emission is highly clumpy and distributed in a
collection of knots with an upper limit on the size of $\sim 0.013$ pc
(limited by the spatial resolution of $0.8\arcsec$)   at a distance
$3.4^{+0.3}_{-0.1}$ kpc \citep{reed}.  The flux densities for the dense
gas knots are listed in Table 1. We selected the brightest knots of 18
regions in the CO, K-continuum or P$\beta$ images  with angular sizes
3$''$--15$''$ and estimated the flux densities after background
subtraction.  The regions include knots which are bright in the
continuum-subtracted CO image and the flux densities of individual
knots range from 0.1 to 0.3  mJy. The flux densities of two
representative CO knots and  one other knot in the NW are given  in
Table 1 along with the positions.  The total flux density of the
CO-detected area after K-continuum subtraction (see Fig. 2a and 2c), is
1.27$\pm$0.17 mJy, where the estimated error includes calibration (7\%),
background variation (5\%) and statistical (10\%) errors. Some faint,
diffuse CO emission from  the Cas A may have been removed from our image
because  of the continuum subtraction; the flux reported here is
conservative, including only the bright knots. After extinction
correction using A$_{v}$ of 4.5 mag \citep{fesen96} and the extinction
law of \cite{rieke85}, the flux density is 2.03$\pm$0.27 mJy, equivalent
to a CO overtone luminosity of 0.01$\pm$0.0014 d$_{3.4kpc}$ $\rm
L_{\odot}$. The ratio of H-continuum to K-continuum  is globally
consistent with the origin of synchrotron radiation with a synchrotron
index of 0.65 - 0.71 (Rho et al. 2003); moreover, the H-continuum and
K-continuum images are similar to the radio image \citep{wright, rho03},
indicating they are dominated by synchrotron emission.
  
We compiled spectral energy distributions (SEDs) of near-infrared H and
K bands and \spitzer\ IRAC Bands 1 to 3 (3.6 - 5.8 \mic) bands for 18
knots;  Figure \ref{casacosed} shows three representative SEDs after
correcting for the extinction. The SEDs show a sudden jump at 2.294
\mic\ toward the CO detection positions, with no such jump toward the
NW ejecta position.  The SEDs of the CO knots also show a larger IRAC
Band  2 (4.5 \mic) excess than that of the NW ejecta knot.

A few knots towards P$\beta$ emitting regions (blue in Figure 1) show
K-continuum brighter than the CO-band emission. These knots  look like
elliptical spheroids, indicative of Quasi Stationary Flocculi (Fesen \&
Gunderson 1996, Willingale et al. 2002). We further note that
H-continuum emission towards hydrogen knots  is slightly  higher than
that expected from synchrotron emission, suggesting  that despite the
narrow bandwidth of H-continuum filter,  the image may be picking up
the weak hydrogen line Br$\lambda$ at 1.57 \mic. The CO filter does not
include any hydrogen or helium lines. The near-infrared spectra of
\cite{gerardy01} did not detect any hydrogen and helium lines within
the H-cont, K-cont, CO filters, but our images have a higher
sensitivity and therefore may be detecting such lines. 

The P$\beta$ emission is largely from shocked circumstellar gas, and
similar to the optical H$\alpha$ emission \citep{fesen96, fesen01}. The
emission is mainly clustered in the northern-most inner shell   and
outside the southwestern shell, as shown in Figure
\ref{casacothreecolor}. The former contains both circumstellar
gas and fast moving knots. The hydrogen lines of circumstellar
medium knots also show moderate high velocity components, which are
defined as $``$mixed emitting knots$"$ by \cite{fesen96}, who discussed
their origin. In particular, rich infrared ejecta emission is detected
from the northern most inner shell  (see Figure 2 of Rho et al. 2008). 
\cite{fesen96} suggested that mixed emitting knots may be $``$splatter$"$
fragments produced as high speed underlying core condensations broke
through and partially mixed with the progenitor star's surface layers. 

\section{Discussion} 
 
We detected significant near-infrared CO overtone emission  from small
portions of the Cas A remnant towards the northern part  of the bright
ring and in the eastern region of Minkowski knots in the  southeast. It
is possible that we are missing fainter CO emission  elsewhere in the
SNR due to our detection limit  (of 1.98$\times$10$^{-6}$ $\rm erg
~s^{-1}~cm^{-2}~sr^{-1}$) in the continuum-subtracted CO image. The
locations where CO emission is detected coincide with the ejecta at the
reverse shock   \citep{hwang04,fesen96,rho08,yang08}. It indicates that
the CO gas was formed in the ejecta.

\begin{figure}[!h]
\epsscale{0.97} 
\plotone{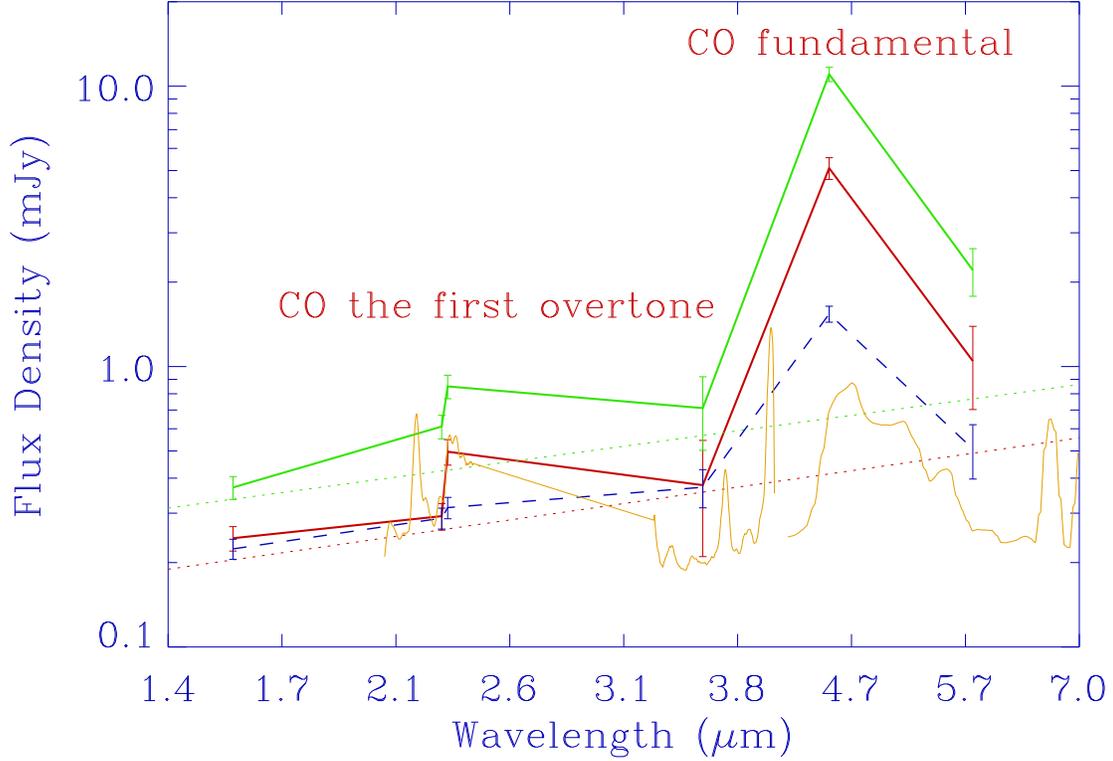}
\caption{The spectral energy distribution of two representative CO
emitting knots: knot1 (green) and knot2 (red) (positions are given in
Table 1)  which show excess emission at 2.294 \mic.   The
contribution from the synchrotron continuum is shown by the dotted
line. Comparison with the SED (scaled to the CO knot1 continuum) of the
NW ejecta knot (blue dashed curve)
shows that  the knots with near-infrared CO-filter excess
emission  may also contain  the CO fundamental emission in the IRAC
band 2.
The error bars of near-infrared emission include both statistical and
calibration errors.
 The SN 1987A spectrum is shown for comparison (orange curve) at
255-260 days after the explosion event (courtesy of Peter Meikle,
Meikle et al. 1989; Wooden et al. 1993).}
\label{casacosed} 
\end{figure}

The \spitzer\ IRAC Band 2 image shows a  morphology similar to
ejecta-dominated images, but the origin of this  emission was a
mystery.  After scaling from the overtone to fundamental band, our
observations  suggest that the CO fundamental contributes some ($>$1/6)
but possibly not  all of the emission in IRAC Band 2 (Figure
\ref{casacosed}). Other  contributors to this band include multiple
[Fe~II], hydrogen  recombination and helium lines. The  ratios of IRAC
2 and CO-filter emission from 18 knots across Cas A  suggest that CO
contributes 10-30\% of the emission in the CO emitting  regions when
compared with non-CO emitting ejecta knots.  Constraining  the
percentage of CO contribution to the IRAC band 2 image is  difficult
since the ratio of the fundamental to the first overtone  varies
depending on the ejecta temperature and density.  We cannot  therefore
strongly constrain the CO contribution to the IRAC Band  2. We analyzed 
the IRS spectroscopy cube to find any signature of the  fundamental
CO (4.8 - 6 \mic), but the signal-to-noise was too poor.  The \spitzer\
IRS data are not  sensitive enough to detect this emission (errors of
the IRS spectra were 1-2 MJy sr$^{-1}$, an order of magnitude higher
than the typical flux densities of near-infrared CO emission)  although there
might be some hint for the northern position (at the position of the
strongest  21 \mic\ dust spectrum).  We also looked for SiO molecules
(at 8 - 10 \mic), but could not distinguish them from the silicate
emission feature  at 9.8 \mic\ (silicate stretching mode).  

We estimated the mass of CO in Cas A by comparing the CO flux density
in Cas A with that found in the supernova SN 1987A, since without
spectra it is not possible to derive a density, temperature, and
velocity of CO gas. Figure \ref{casacosed} compares the CO emission
from Cas A and  SN 1987A \citep{meikle89,wooden93}. After accounting
for the distance differences,  the CO (2-0) first overtone flux density
of Cas A is a factor of $\sim 28$ smaller  than that of SN1987A on day
255.   Assuming the same gaseous conditions as seen in the CO emitting
regions in SN1987A (a temperature of 1800-2800 K and a velocity of 2000
km~s$^{-1}$), we estimated the CO mass in Cas A to be  $\rm M_{CO}\sim
1.7\times 10^{-6} \,\rm M_{\odot}$  for Local Thermodynamic 
Equilibrium (LTE) \citep{spyromilio88} and $8.2\times 10^{-6} \rm
\,M_{\odot}$ for non-LTE  \citep{liu92}. Note that the CO mass estimate
is highly uncertain because we assume the gas properties described
above.

 The CO detections in SNe have been within a few years after the
initial explosion.  The main destruction process of CO is from impacts
with energetic  electrons \citep{liu92} and with ionized gas atoms. 
The latter  process quickly destroys CO, hence the detection of the CO
fundamental  band provides constraints on the degree of mixing in the
SN ejecta.  Given that Cas A is believed to be the remnant of a  Type
IIb explosion (Krause et al. 2008), this suggests that either the  CO
layer is not macroscopically mixed with ionised helium or that the 
helium is not ionised.   Typical electron density of C- and O-rich
regions during formation of CO is shown to be 5$\times$10$^8$ cm$^{-3}$
\citep{gearhart99}.  The highly clumpy structures can be explained by
such high density CO gas within the warm or hot  ejecta,  
which naturally fortifies the evidence of fresh dust formation
observed in Cas A \citep{lagage96, rho08}.

In dust evolution models,  understanding CO molecules in SN ejecta is a
key physical question.  The formation efficiency  of CO molecules
critically affects the dust formation efficiency.  In chemical
equilibrium, the abundance of CO is balanced between radiative
processes and radioactive decay fo $^{56}$Co, because the main
destruction process is by the energetic electrons produced by the
radioactive decay of $^{56}$Co \citep{liu92,clayton01}. Both Nozawa et
al. (2003) and \cite{todini01}  assumed the mass  fraction of C atoms
that are not locked in CO is 99-100\%, and \cite{travaglio99} assumed
that mixing occurs   at the molecular level, prior to dust
condensation. In contrast, Clayton \& Nittler (2004), Deneault,
Clayton \& Heger (2003), and \cite{fryxell91} suggest that gases are
unlikely to be mixed at the molecular level within a few years. Our CO
detection suggests that molecular mixing is small (at least smaller
than previously thought) in the SN ejecta and during the development
of the reverse shock. CO molecules in Cas A suggests a higher
CO abundance in the gas phase and lower efficiency in dust formation
of carbon dust. Remaining molecules also indicate that dust formation
was inefficient (leaving molecular leftovers) or occurred
inhomogeneously  (e.g. only in dense knots).  Deep
infrared spectroscopic observations  
will allow us to measure the velocity, temperature and
density of the CO molecules in Cas A and to study the on-going
chemistry of molecules in SN ejecta.

\section{Conclusions}
We detected carbon monoxide emission from the young supernova remnant
Cas A.  This may be the first CO detection from an older SN (age $\sim$300
yr).  CO emission has also been detected in 8 other core-collapse
remnants, but in all cases it has been limited to within 3 years of
the explosion.  Theoretical models suggest that dust formation in  SNe
occurs between 200-800 days after the SN event yet the  observations
presented here suggest that molecular and dust formation  processes
could continue long after. The detection of CO in the
Cas A remnant demonstrates  that astrochemical processes are still
on-going more than 300 years after the initial explosion.   
  
\acknowledgements We are grateful to Achim Tappe who participated in
the Palomar observing run, and thank Takashi Onaka for fruitful
discussion on dust and CO formation.  We also thank Palomar observatory staff
(Jean Mueller, Jeff Hickey and Karl Dunscombe) for assistance during
our observing runs.  
We thank the anonymous referee for careful reading and helpful comments.
H. Gomez would like to thank Las Cumbres
Observatory for their support. 
M. Andersen is partially supported by NASA through LTSA grant
NRA-01-01-LTSA-013.
 Based on observations obtained at the
Hale Telescope, Palomar Observatory as part of a continuing
collaboration between the California Institute of Technology,
NASA/JPL, and Cornell University.

{\scriptsize
\begin{table} 
\caption[]{Summary of Observed Brightnesses} 
\label{Tinter} 
\vspace{-0.3 cm} 
\scriptsize 
\begin{center} 
\begin{tabular}{llllcccccc} 
\hline \hline 
filter & flux density$^a$  &  \multicolumn{3}{c}{Flux Density (mJy)} & 
\\ 
\cline{3-5}
       &         range  & CO knot1$^b$ & CO knot2$^b$ & NW  knot$^b$&   \\ 
\hline 
2.27 $\mu$m (K-cont)&  0.1-1.0      & 0.610$\pm$0.009 &0.290$\pm$0.006 &1.780$\pm$0.016\\ 
2.29 $\mu$m (CO) &0.2-1.2            & 0.850$\pm$0.013 & 0.490$\pm$0.010 & 1.880$\pm$0.020\ \\ 
3.6 $\mu$m (IRAC 1)     & 0.4-2.3   & 0.71$\pm$0.20 & 0.38$\pm$0.16 & 2.23$\pm$0.35 &  \\ 
4.5 $\mu$m (IRAC 2) & 1.2 -15.0  & 11.03$\pm$0.64 & 5.10$\pm$0.45 & 9.24$\pm$0.60 &  \\ 
5.8 $\mu$m  (IRAC 3) & 0.6-4.2 &  2.20$\pm$0.42 & 1.04$\pm$0.34 & 3.05$\pm$0.67  &\\ 
\hline 
\end{tabular} 
\end{center} 
$^{a}${The range of flux densities (in mJy) towards CO detected positions. 
$^b$Positional coordinates in RA and Dec (J2000): CO knot1 (green in
Figure \ref{casacosed}) at  350.850$^{\circ}$, 58.840$^{\circ}$, knot2
(red) at 350.854$^{\circ}$, 58.834$^{\circ}$, and  the northwest ejecta
knot (blue) at 350.825$^{\circ}$ and 58.836$^{\circ}$. The flux densities were 
measured by averaging over an angular size of 5.3$''\times$3.1$''$,
38$''\times$10$''$, and  6$''\times$12$''$ for knot1, knot2, and the NW
knot, respectively. } 
\end{table} 
}

{}

\end{document}